\documentclass[aps,prb,twocolumn,groupedaddress]{revtex4-1}

\usepackage{graphicx}
\usepackage[english]{babel}
\usepackage{enumitem}
\usepackage{multirow}
\usepackage{array}
\newcolumntype{N}{>{\centering\arraybackslash}m{2cm}}
\newcolumntype{G}{>{\centering\arraybackslash}m{3cm}}

\begin{document}


\title{CSAR 62 as negative-tone resist for high-contrast e-beam lithography at temperatures between 4~K and room temperature}


\author{Arsenty Kaganskiy}
\email[]{arsenty.kaganskiy@tu-berlin.de}
\affiliation{Institut f{\"u}r Festk{\"o}rperphysik, Technische Universit{\"a}t Berlin, Hardenbergstra{\ss}e 36, D-10623 Berlin, Germany}

\author{Tobias Heuser}
\affiliation{Institut f{\"u}r Festk{\"o}rperphysik, Technische Universit{\"a}t Berlin, Hardenbergstra{\ss}e 36, D-10623 Berlin, Germany}

\author{Ronny Schmidt}
\affiliation{Institut f{\"u}r Festk{\"o}rperphysik, Technische Universit{\"a}t Berlin, Hardenbergstra{\ss}e 36, D-10623 Berlin, Germany}

\author{Sven Rodt}
\affiliation{Institut f{\"u}r Festk{\"o}rperphysik, Technische Universit{\"a}t Berlin, Hardenbergstra{\ss}e 36, D-10623 Berlin, Germany}

\author{Stephan Reitzenstein}
\affiliation{Institut f{\"u}r Festk{\"o}rperphysik, Technische Universit{\"a}t Berlin, Hardenbergstra{\ss}e 36, D-10623 Berlin, Germany}


\date{\today}

\clearpage

\begin{abstract}
The temperature dependence of the electron-beam sensitive resist CSAR~62 is investigated in its negative-tone regime. The writing temperatures span a wide range from 4~K to room temperature with the focus on the liquid helium temperature regime. The importance of low temperature studies is motivated by the application of CSAR~62 for deterministic nanophotonic device processing by means of in-situ electron-beam lithography. At low temperature, CSAR~62 exhibits a high contrast of 10.5 and a resolution of 49~nm. The etch stability is almost temperature independent and it is found that CSAR 62 does not suffer from peeling which limits the low temperature application of the standard electron-beam resist PMMA. As such, CSAR~62 is a very promising negative-tone resist for in-situ electron-beam lithography of high quality nanostructures at low temperature.
\end{abstract}


\maketitle

\section{Introduction}
The development of functional quantum devices based on single quantum dots is an emerging field in semiconductor nanotechnology \citep{S.Buckley.2012, P.Lodahl.2013}. One of the big challenges is the precise integration of a single quantum emitter into a nanophotonic device structure \citep{S.Reitzenstein.2012}. To overcome this issue, deterministic lithography techniques were developed that combine the tasks of spectroscopic identification of a single high quality emitter with instant lithography. This was realized by a combination of optical spectroscopy and optical lithography \cite{Dousse.2008} and by combined cathodoluminescence spectroscopy (CL) and electron-beam lithography (EBL) \cite{Gschrey.2013}, dubbed CLL. Due to the emission properties of most semiconductor quantum dots, such processing has to be performed at low cryogenic temperatures or even liquid helium (l-He) temperature for sufficient signal-to-noise ratio \citep{M.Bayer.2002}. Naturally, EBL is the most attractive technique to fabricate very precise or even three dimensional device structures with flexible design\citep{Y.Chen.2015, V.Kudryashov.2003, W.H.Teh.2003, Schnauber.2016}. This initiates the quest for a detailed understanding of low-temperature-capable electron-beam-sensitive resists \citep{M.Gschrey.2014}. Additionally, a resist with a low sensitivity in the negative-tone regime is needed for CLL: the preceding spectroscopic investigation on the resist-coated sample requires dwell times per pixel of 10 to 30 ms. The corresponding electron dose must not lead to an effective exposure of the resist. This points to EBL resists that can be operated in both a positive-tone (via chain-scission at small doses) and negative-tone (via cross-linking at large doses) regime, with the effective doses for the negative-tone regime being orders of magnitude larger than for the positive-tone regime. Negative-tone resists also exhibit a higher mechanical stability of written structures due to the brunched chain structure that results from the cross-linking process \citep{W.M.Yeh.2010}. A prominent example for this classes of resists is polymethyl methacrylate (PMMA). Negative-tone PMMA was already applied for, e.g., the fabrication of quasi-three-dimensional micro/nanomechanical components \citep{W.H.Teh.2003} and planar cobalt electrodes separated by a sub-10~nm gap \citep{L.Ressier.2007}. Also in its negative-tone regime it acts as a high-resolution resist \citep{I.Zailer.1996, A.C.F.Hoole.1997}. It was found to be suitable for low-temperature CLL \citep{Kaganskiy.2015, Gschrey.2013}, especially for 3D patterning \citep{Schnauber.2016, Gschrey.2015}. A drawback in its use, however, is the necessity of a two-step lithography approach to gain a high device yield \citep{Schnauber.2016}. Also a higher contrast is desirable for the fabrication of filigree structures with steep facets like ring resonators \citep{M.Davanco.2011} and photonic crystals \citep{T.D.Happ.2002}. Besides PMMA and ZEP (see below) there are some more positive-tone/negative-tone resists like metal-carbonyl organometallic polymers \citep{J.Zhang.2015}, poly(methylglutarimide) \citep{G.Karbasian.2012}, phenyl-bridged polysilsesquioxane \citep{L.Brigo.2012}, and triphenylene derivatives \citep{A.P.G.Robinson.1999}.

Here we present investigations on the EBL resist AR-P~6200 (CSAR~62) from Allresist GmbH in its negative-tone regime and in the full range between l-He temperature and room temperature. It is a poly-$\alpha$-methylstyrene-co-methyl-chloroacrylate based resist together with an acid generator dissolved in anisole that was originally designed to be used as a positive-tone resist. However, it is similar to the positive-tone resist ZEP that was shown to be suitable for negative-tone lithography also \citep{Oyama.2011, M.A.Mohammad.2012}. We demonstrate that CSAR is very suitable for the aforementioned requirements of deterministic low-temperature CLL without the need for a two-step exposure.

\section{Methods}
EBL at various temperatures was performed in a setup based on a Jeol JSM~840 scanning electron microscope (SEM) which was extended by a custom-made low-temperature lithography attachment including a liquid helium flow cryostat. It also features an extension for cathodoluminescence spectroscopy\cite{Gschrey.2015b}. The commercial resist under investigation is AR-P~6200 (CSAR) in anisole solvent (4~$\%$ and 9~$\%$ solid content for resist thicknesses of 78 and 210~nm, respectively). Prior to coating, the used underlying GaAs substrate was cleaned for 5~minutes in acetone and isopropanol at 70$^\circ$C, respectively. Two resist thicknesses of 78~nm and 210~nm were fabricated by spin coating  with a rotation frequency of 6000 and 3000 rotations per minute, respectively, as additional process parameter. After that, the samples were post-baked for 1 minute at 175$^\circ$C. The resist thickness was determined by mechanical removing of the resist and subsequent profilometry across the area with and without the resist. The uncertainty of the measurement was $\pm$~3~nm. Motivated by the upcoming low-temperature lithography techniques, we studied the properties of CSAR~62 in the temperature range between 4~K and 295~K. EBL was done with an acceleration voltage of 20~kV and beam currents of 0.5~nA and 1~nA. In order to study the onset-dose and the contrast of CSAR~62 in the negative-tone regime, contrast curves with a linear dose gradient were written. Each dose gradient is made up of a 75~$\mu$m long and 15~$\mu$m broad stripe, which consists of equally spaced lines with a 10~nm step width (corresponding to $\approx$~6.7~$\mu$C/cm$^2$ dose step width) written with a linearly increased nominal dose from 0 to 50~mC/cm$^2$. The developed and etched structures were analyzed with the profilometer Ambios XP2 (160 nm tip step width). By using marker lines above and under the gradient stripe the scan length measured by profilometer was converted into the written dose (Fig. \ref{fig:Contrast-Curves} (a)).

To adapt the development process to the low-temperature writing, different development recipes were applied (see next section). Best results were obtained after dip developing for 45~s with the commercial developer AR~600-546 by Allresist. Then the samples were dipped for 30~s in isopropanol (Isopropanol Micropur by Technic France, VLSI purification grade) followed by a 30~s dip in deionized water to stop the process.  In order to analyze the etch selectivity of the inverted resist the samples were dry etched using inductively-coupled-plasma reactive-ion etching (ICP-RIE). As process gases for the plasma 1.3~sccm Cl$_2$, 4.3~sccm BCl$_3$ and 1.1~sccm Ar were used. The etching was done with a pressure of 0.08~Pa, CCP power of 24.5~W, ICP power of 100~W and bias voltage of -215~V. 

Arrays of lines and dots with varying widths and diameters were written and processed to determine the resolution of the resist. After the resist development as well as resist etching they were evaluated with a Digital Instruments Multimode AFM (MMAFM-2) in a constant force contact mode and with a Zeiss Ultra 55 SEM using a low acceleration voltage of 0.8~kV to avoid charging artifacts and achieve a charge balance regime\citep{Wuhrer.2016}.

\section{Results and Discussion}

First we optimized the development process of the inverted low-temperature resist by comparing results for different developers (AR~600-546, AR~600-548, and AR~600-56 by Allresist, isopropanol (IPA) : water 7~:~3 solution and pure IPA) in combination with various development times. All samples (except of those which were developed in IPA or in IPA : water solution) were dipped after the development for 30~s in IPA and swiveled for 30~s in deionized water. 

\begin{figure}[t!]
\centering
\includegraphics[width=8 cm]{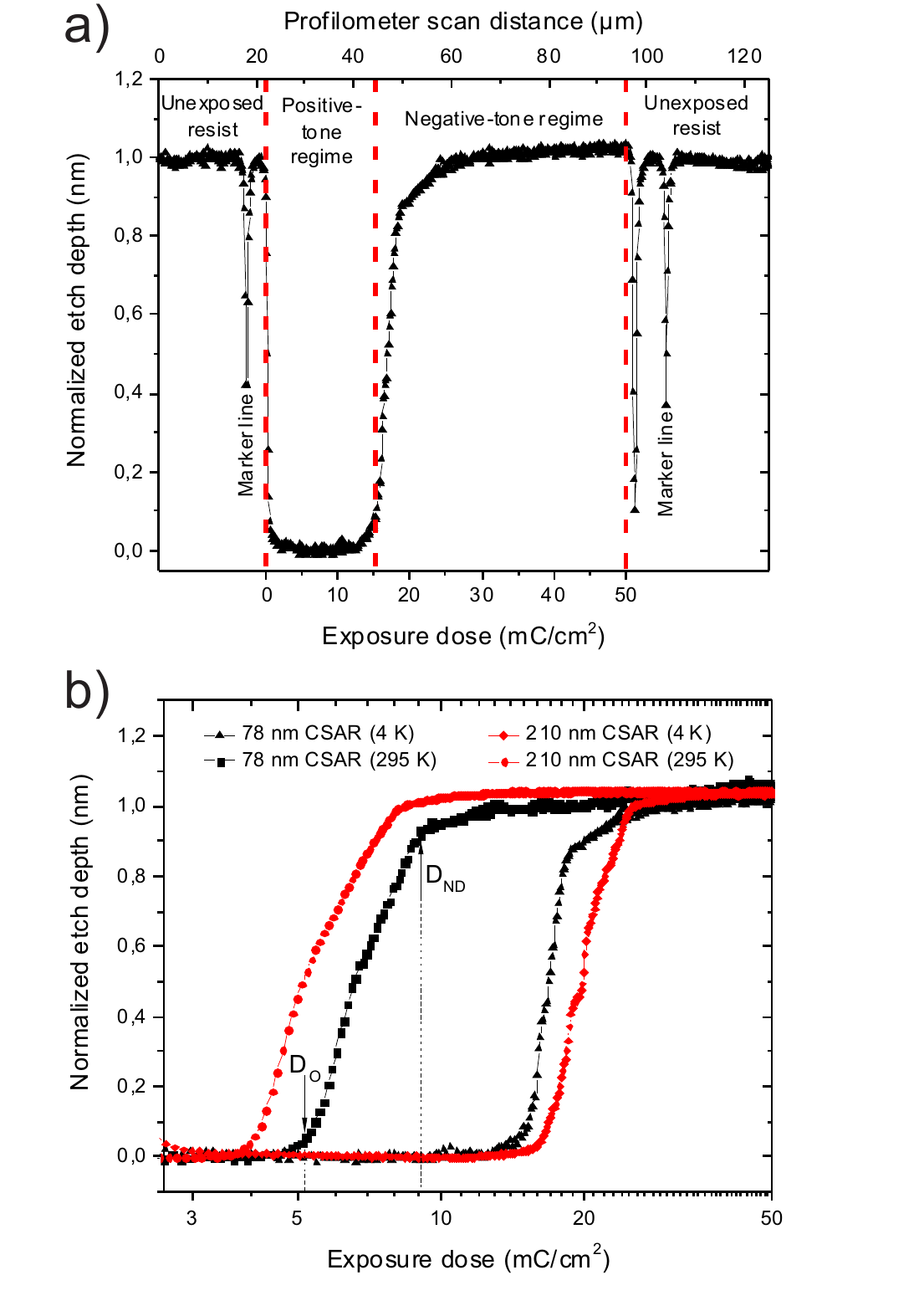}
\caption{(Color Online) (a) Etched contrast curve of 78~nm thick CSAR~62 written at 4~K. The curve is plotted on a linear scale and schematically divided into unexposed, positive-tone and negative-tone regions. The marker lines used for the accurate assignment of electron dose to lateral position are shown. (b) Etched contrast curves for 78~nm (black) and 210~nm (red) thick CSAR~62 written at 4 and 295~K, respectively. The curves are plotted on a half logarithmic scale. All curves are restricted to the negative-tone regime, where the resist starts to cross-link. The slopes of the curves refer to the contrast, which is evaluated by equation~\ref{eq:Contrast} and given in table \ref{Table-Contrast}. Doses $D_{O}$ and $D_{ND}$ are shown exemplary for a 78 nm thick CSAR written at 295~K.}
\label{fig:Contrast-Curves}
\end{figure}

At first we used the developer AR~600-546 (based on amyl acetate) and developed a sample for 45~s. This time span was enough to complete the development process and it caused no resist peeling or structural damage as observed e.g. for low-temperature-exposed PMMA\citep{Schnauber.2016}. Due to the fact that the developer AR~600-548 (main component diethyl ketone~/~diethyl malonate) is more aggressive than AR~600-546 and causes dark erosion, we started from a 5~s development, which was not sufficient to develop tight trenches in the resist. Another 5~s of development lead to a peeling of the resist on the entire sample surface. So we skipped further tests on that developer. Next we developed a sample for 5~s with AR~600-56 (on the basis of methyl isobutyl ketone (MIBK)) resulting in a complete development, being comparable to the development for 45~s in AR~600-546. Finally, we tested IPA and IPA~:~water solution. Neither positive-tone nor negative-tone CSAR~62 were completely developed after the developing for 10~s in pure IPA and for 20~s in a 7~:~3 solution of IPA~:~water (77~$\%$ clearance in the positive-tone regime). Another drawback is a reduced onset dose ($D_O$) of 8.5~mC/cm$^{2}$ (3.1~mC/cm$^{2}$) for the 78~nm thick resist developed in pure IPA (IPA-water solution) in contrast to 15~mC/cm$^{2}$ for the sample developed with AR~600-546. In-situ spectroscopic tasks via CL are limited by the time window before cross-linking starts at $D_O$. Both onset doses hinder such investigations. Together with a reduced sensitivity in the positive regime it confirms previous results\citep{Koshelev.2011, M.A.Mohammad.2012}. In conclusion, we decided to develop the investigated samples for 45~s with the developer AR~600-546.

\begin{figure}[t!]
\centering
\includegraphics[width=8 cm]{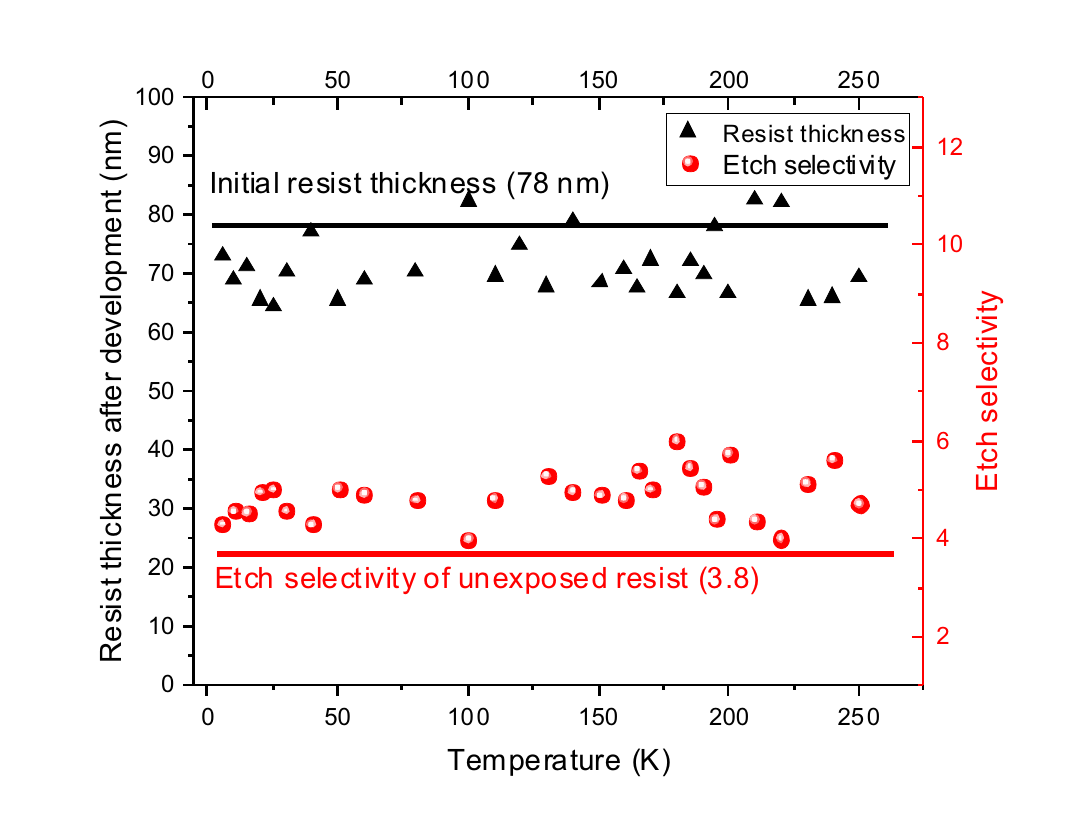}
\caption{(Color Online) Remaining resist thickness of formerly 78~nm thick CSAR~62 after the development (black). The thickness reduction of 8~$\%$ corresponds to a moderate outgassing of HCl and other compounds. Etch selectivity of negative-tone CSAR~62 (red). In all cases the applied electron dose was 40~mC/cm$^{2}$. The thickness and etch selectivity are almost constant in the whole temperature range.}
\label{fig:Etching}
\end{figure}

To accurately evaluate the contrast curves and to obtain the contrast and the onset dose, etching into the underlying GaAs was performed for two reasons: First, the resist gets scratched by the needle of the profilometer which distorts the results and, second,  we directly obtain the etched profiles which is of large interest for the envisaged transfer of resist patterns into the semiconductor. The only direct profilometry on the resist was performed to get the resist thickness of fully inverted CSAR~62 (cf.~Fig.~\ref{fig:Etching}). Figure \ref{fig:Contrast-Curves} (a) presents an example of an etched contrast curve that was enclosed into regions of unexposed resist. As mentioned before, at small exposure doses the resist can be completely removed from the sample surface during the development (positive-tone regime) while at higher doses its components cross-link and the resist becomes insoluble for the developer (negative-tone regime). Figure~\ref{fig:Contrast-Curves} (b) displays etched dose gradients on a logarithmic scale zoomed into the negative-tone regime. They were written at different temperatures into resist layers with a thickness of 78~nm and 210~nm, respectively. The etching was stopped at the point when the unexposed resist was completely removed. The etch time was 115~s (250~s) for the 78~nm (210~nm) thick resist. The etch depth is normalized with respect to the regions of unexposed resist. The black (red) curves correspond to an initial resist thickness of 78~nm (210~nm) and writing temperatures of 295~K and 4~K were exploited. 

In order to calculate the contrast $\gamma$ of the resist in the negative-tone regime we used the common equation:
\begin{eqnarray}
\gamma = -\frac{1}{\log(D_{O}/D_{ND})}
\label{eq:Contrast}
\end{eqnarray}
$D_{O}$ is the onset dose for the cross linking and marks the begin of the curves' upward slopes. $D_{ND}$ gives the dose for which the resist is fully cross-linked and the etch depth starts to be constant. Consequently, the contrast is directly expressed by the slope of the curve between $D_{O}$ and $D_{ND}$. Table~\ref{Table-Contrast} summarizes the contrast values as extracted from Fig.~\ref{fig:Contrast-Curves}.

\begin{figure}[t!]
\centering
\includegraphics[width=8 cm]{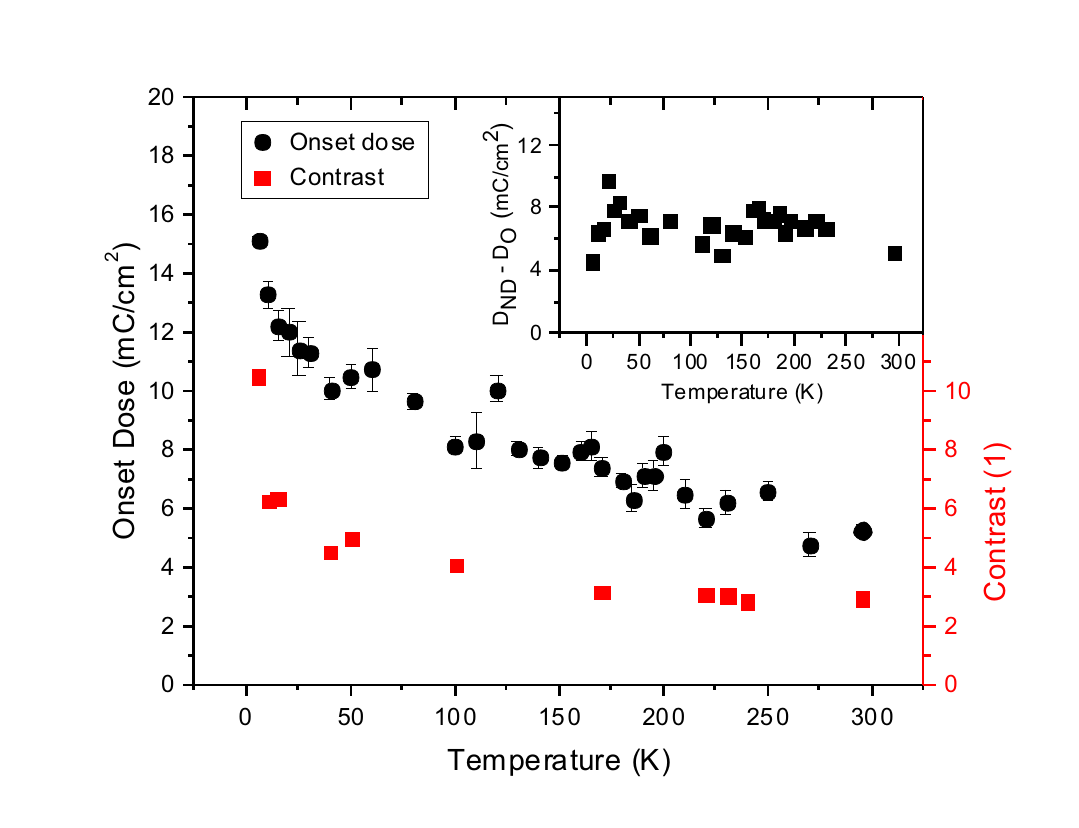}
\caption{(Color Online) Onset dose (black) and contrast (red) of 78~nm CSAR~62 as a function of the writing temperature. The onset-dose decreases with increasing temperature due to accelerated cross-linking of free radicals. The inset shows the temperature dependence of the additional electron dose that is required to finish the cross-linking process when it has started at $D_{O}$.}
\label{fig:t-dependence}
\end{figure}

\begin{table}[bb]
\caption{Contrast of CSAR~62 for two different temperatures and resist thicknesses. The highest value of 10.5 corresponds to a resist thickness of 78~nm and a writing temperature of 4~K.\\}
{\renewcommand{\arraystretch}{1.3}
\begin{tabular}{GNN}
\hline \hline
\multicolumn{1}{G}{Temperature} & \multicolumn{2}{c}{Resist thickness} \\ \hline
\multicolumn{1}{c}{}            & 78 nm   & 210 nm \\  \cline{2-3}
\multicolumn{1}{c}{4 K}         & 10.5    & 5.7     \\ 
\multicolumn{1}{c}{295 K}       & 2.97    & 2.84     \\ \hline \hline
\end{tabular}}
\label{Table-Contrast}
\end{table}

The highest contrast of 10.5 is observed for the resist thickness of 78~nm and l-He temperature.  A higher resist thickness results in a reduced contrast as does a higher temperature. It is worth to mention that the contrast value of 5.7 measured for a 210~nm thick CSAR~62 written at 4~K is almost 5 times higher than the contrast of a similar thick PMMA written at the same temperature as reported by us previously\citep{Schnauber.2016}. 

Due to its higher contrast and resolution and hence its wider potential application spectrum the 78~nm thick resist was subsequently investigated in detail by evaluating etched contrast curves in the full temperature range between 4~K and 295~K. Fig.~\ref{fig:t-dependence} (black circles) shows the onset dose of the negative-tone regime as a function of temperature. Clearly, the onset dose decreases monotonously with increasing temperature. The processes occurring during the transition from positive- to negative-tone regime in a similar ZEP resist were studied by Oyama et al.\citep{Oyama.2012} They showed that a driving force leading to a main chain scission under e-beam exposure is a dissociation of the chlorine atom of the $\alpha$-chloromethacrylate unit of the chain via dissociative electron attachment, which leads to the formation of temporal double bonds. Thereby a number of Cl bonds and amounts of Cl decrease rapidly for an increasing exposure dose. The cross-linking occurs during a further exposure at higher doses, at which free radicals are activated at the cost of a reduced number of double bonds and an increasing number of C-H bounds\citep{Oyama.2011, Oyama.2014}. Due to a higher mobility of activated and free radicals at higher writing temperatures the cross-linking starts at smaller onset doses (Fig. \ref{fig:t-dependence}). 

A similar trend is observed for the contrast (red squares in Fig.~\ref{fig:t-dependence}). To asses this behavior it is necessary to evaluate both doses that are considered in equation~\ref{eq:Contrast}. As the cross linking starts from the onset dose, the dose difference $D_{ND}$-$D_{O}$ indicates the electron dose that has to be additionally deposited to fully invert the resist. As it can be seen in the inset of Fig.~\ref{fig:t-dependence}, $D_{ND}$-$D_{O}$ is almost temperature independent, which also reflected in the identical slopes of linearly plotted contrast curves (not shown here). Together with the trend of $D_{O}$ and the logarithmic definition of the contrast value (Eqn.~\ref{eq:Contrast}) this results in the observed temperature dependence of the contrast. A similar behavior of decreasing contrast with increasing temperature was observed by Sidorkin et al. \citep{Sidorkin.2008} for hydrogen silsesquioxane (HSQ), however no explanation could be given.

The thickness of fully inverted resist after its development was also investigated as a function of temperature (black triangles in Fig.~\ref{fig:Etching}). It is independent on temperature and has decreased on average by about 6.3~nm ($8 \%$) as compared to the initial resist thickness of 78~nm. This is attributed to the elimination of Cl from the resist and subsequent outgassing of HCl and other scission products and the buildup of a more dense cross-linked compound \citep{Oyama.2012}. Additionally and consequently, also the etch selectivity against GaAs is almost constant over the whole temperature range with an increase of $\approx 30\%$ relative to the unexposed resist. This is an ideal situation for deterministic nanoprocessing as it enables constant etching and processing conditions over the entire range of writing temperatures. Additionally, the etch selectivity of $\approx 4.9$ in combination with an increased etch stability of CSAR~62 as compared to PMMA ($\approx$ by a factor of 2) allows for direct etching without the need for intermediate hard masks. 

\begin{figure}[t!]
\centering
\includegraphics[width=8 cm]{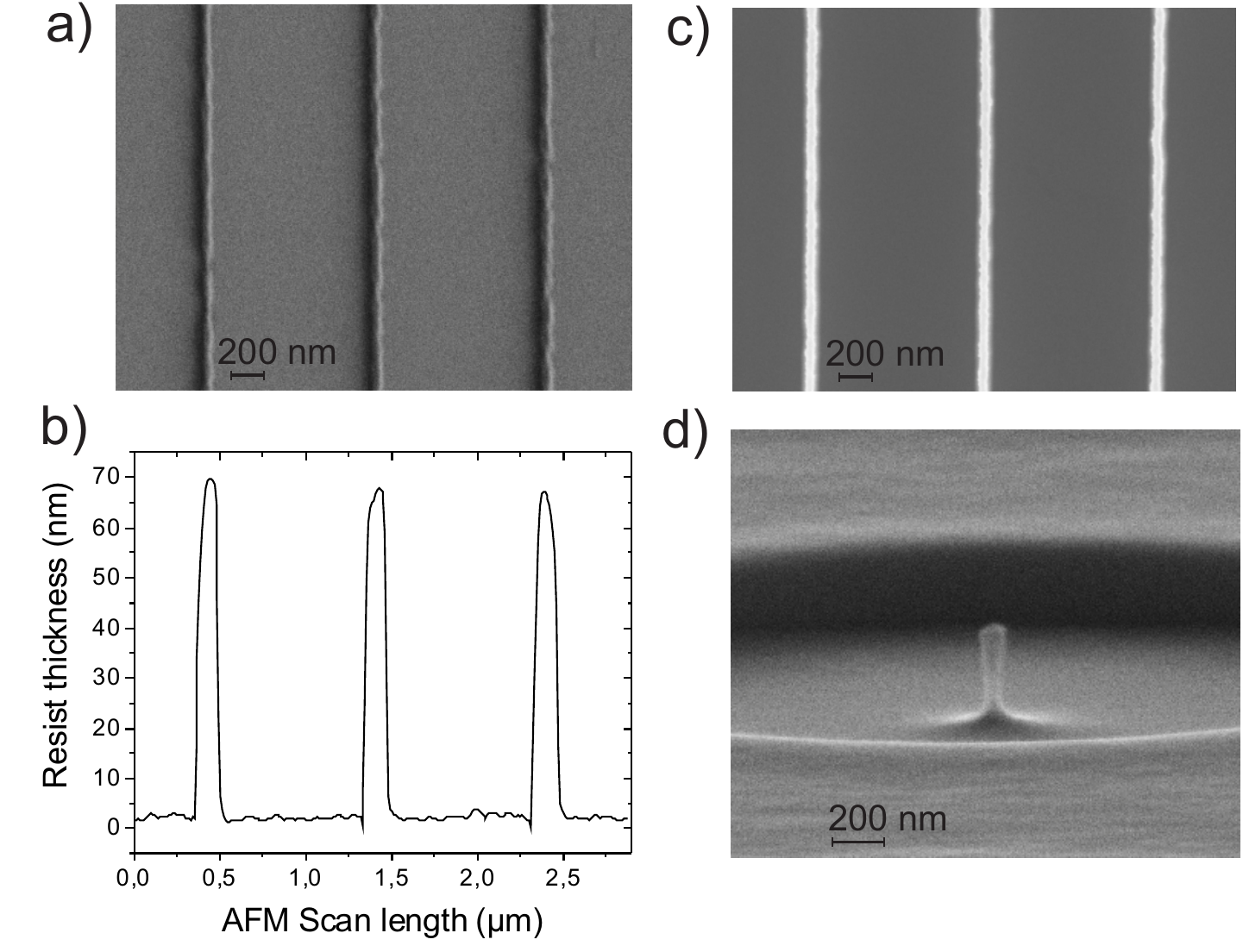}
\caption{(Color Online) SEM images of developed (a)  and etched (c) line arrays taken using a secondary electron detector and an in-lens detector, respectively. The width of the lines is 49~nm. (b) AFM profile of a group of narrow lines after the development. (d) Etched pillar structure with a diameter of 49~nm. All structures were written into 78~nm thick CSAR~62 in its negative-tone regime and at a temperature of 4~K.}
\label{fig:REM}
\end{figure}

These results for the resist thickness and etch selectivity are in contrast to our experimental findings on PMMA. Low-temperature negative-tone PMMA was found to suffer from an extensive swelling due to trapped gaseous scission products around 70~K that also reduced the etch selectivity in this temperature regime in a drastic way \citep{M.Gschrey.2014}. Low-temperature writing of PMMA also requires a two-step lithography technique to prevent it from peeling when trapped gases burst out \citep{Unai.2013, Schnauber.2016}. In the case of CSAR~62 we did not observe such structural damage for structures written at low-temperatures, which makes it much more suitable for low temperature EBL. It is noteworthy that the reproducibility is very good by means of characteristic doses and quality of the written structures.

In order to determine the resolution capability of low-temperature CSAR~62 we performed lithography of line arrays and dots at 4~K written into a 78~nm thick resist layer. The lines were written with a pitch of 1~$\mu$m. They consist of aligned equidistant dots while their width was varied by means of an increasing exposure dose. We could observe continuous lines starting from a width of 49~nm as it can be seen in the SEM image in Fig.~\ref{fig:REM}~(a).
The lines are well resolved, structurally intact and exhibit low line width and edge roughness. In Fig.~\ref{fig:REM}~(b) an AFM scan of such narrow lines is shown. Due to the aforementioned high contrast of CSAR~62 one can see pronounced steep edges of the lines. A slight broadening of the lines is caused by convolution with the AFM tip. SEM image of a group of 49~nm wide lines after etching is shown in Fig.~\ref{fig:REM}~(c). It could be clearly seen that the high resolution and edge steepness of the resist could be completely transferred in the etched GaAs. The aforementioned high resolution of the resist is confirmed by a number of pillars written in the positive-tone as well as in the negative-tone regime which exhibit the smallest diameter of the developed pillars of 48 and 49~nm, respectively. An etched pillar with a diameter of 49~nm written in the negative-tone regime is exemplary shown in Fig.~\ref{fig:REM}~(d). In conclusion, we obtain a resolution that is even better than that of low-temperature negative-tone PMMA, which exhibited at best a resolution of 65~nm \citep{Gschrey.2015b}.

From AFM measurements we evaluated the average surface roughness and root mean square roughness of 0.43~nm and 0.58~nm, respectively, measured over an area of 5~x~5~$\mu$m$^2$ of the negative-tone (40 mC/cm$^2$) 78~nm thick resist at 4~K after the development.

We also want to point out our finding that negative-tone CSAR~62 can be easily removed in a moderate oxygen plasma, while the surface of the semiconductor sample is not damaged. This feature is crucial for the residue-free processing of nanophotonic devices.

\section{Summary and Conclusion}

The EBL resist CSAR~62 has been investigated in the temperature range between 4~K and room temperature in its negative-tone regime. At low temperatures it exhibits high contrasts of up to 10.5 that are linked to the temperature dependence of the onset dose. The resist thickness and etch selectivity are almost independent on temperature and no detrimental swelling or peeling of low-temperature written structures was observed. The resolution at l-He temperature was better than 50~nm. This makes CSAR~62 a very attractive resist for novel low-temperature applications in the field of deterministic semiconductor device technology. 
\\

\section{Acknowledgments}

The research leading to these results has received funding from the European Research Council under the European Union's Seventh Framework ERC Grant Agreement No. 615613, and from the Deutsche Forschungsgemeinschaft (DFG) through SFB 787 $"$Semiconductor Nanophotonics: Materials, Models, Devices$"$.


\begin{thebibliography}{32}%
\makeatletter
\providecommand \@ifxundefined [1]{%
 \@ifx{#1\undefined}
}%
\providecommand \@ifnum [1]{%
 \ifnum #1\expandafter \@firstoftwo
 \else \expandafter \@secondoftwo
 \fi
}%
\providecommand \@ifx [1]{%
 \ifx #1\expandafter \@firstoftwo
 \else \expandafter \@secondoftwo
 \fi
}%
\providecommand \natexlab [1]{#1}%
\providecommand \enquote  [1]{``#1''}%
\providecommand \bibnamefont  [1]{#1}%
\providecommand \bibfnamefont [1]{#1}%
\providecommand \citenamefont [1]{#1}%
\providecommand \href@noop [0]{\@secondoftwo}%
\providecommand \href [0]{\begingroup \@sanitize@url \@href}%
\providecommand \@href[1]{\@@startlink{#1}\@@href}%
\providecommand \@@href[1]{\endgroup#1\@@endlink}%
\providecommand \@sanitize@url [0]{\catcode `\\12\catcode `\$12\catcode
  `\&12\catcode `\#12\catcode `\^12\catcode `\_12\catcode `\%12\relax}%
\providecommand \@@startlink[1]{}%
\providecommand \@@endlink[0]{}%
\providecommand \url  [0]{\begingroup\@sanitize@url \@url }%
\providecommand \@url [1]{\endgroup\@href {#1}{\urlprefix }}%
\providecommand \urlprefix  [0]{URL }%
\providecommand \Eprint [0]{\href }%
\providecommand \doibase [0]{http://dx.doi.org/}%
\providecommand \selectlanguage [0]{\@gobble}%
\providecommand \bibinfo  [0]{\@secondoftwo}%
\providecommand \bibfield  [0]{\@secondoftwo}%
\providecommand \translation [1]{[#1]}%
\providecommand \BibitemOpen [0]{}%
\providecommand \bibitemStop [0]{}%
\providecommand \bibitemNoStop [0]{.\EOS\space}%
\providecommand \EOS [0]{\spacefactor3000\relax}%
\providecommand \BibitemShut  [1]{\csname bibitem#1\endcsname}%
\let\auto@bib@innerbib\@empty
\bibitem [{\citenamefont {{S. Buckley}}\ \emph {et~al.}(2012)\citenamefont {{S.
  Buckley}}, \citenamefont {{K. Rivoire}},\ and\ \citenamefont {{J.
  Vuckovic}}}]{S.Buckley.2012}%
  \BibitemOpen
  \bibfield  {author} {\bibinfo {author} {\bibnamefont {{S. Buckley}}},
  \bibinfo {author} {\bibnamefont {{K. Rivoire}}}, \ and\ \bibinfo {author}
  {\bibnamefont {{J. Vuckovic}}},\ }\href {\doibase
  10.1088/0034-4885/75/12/126503} {\bibfield  {journal} {\bibinfo  {journal}
  {Rep. Prog. Phys.}\ }\textbf {\bibinfo {volume} {75}},\ \bibinfo {pages}
  {126503} (\bibinfo {year} {2012})}\BibitemShut {NoStop}%
\bibitem [{\citenamefont {{P. Lodahl}}\ and\ \citenamefont {{S.
  Stobbe}}(2013)}]{P.Lodahl.2013}%
  \BibitemOpen
  \bibfield  {author} {\bibinfo {author} {\bibnamefont {{P. Lodahl}}}\ and\
  \bibinfo {author} {\bibnamefont {{S. Stobbe}}},\ }\href {\doibase
  10.1515/nanoph-2012-0039} {\bibfield  {journal} {\bibinfo  {journal}
  {Nanophotonics}\ }\textbf {\bibinfo {volume} {2}},\ \bibinfo {pages} {39}
  (\bibinfo {year} {2013})}\BibitemShut {NoStop}%
\bibitem [{\citenamefont {{S. Reitzenstein}}(2012)}]{S.Reitzenstein.2012}%
  \BibitemOpen
  \bibfield  {author} {\bibinfo {author} {\bibnamefont {{S. Reitzenstein}}},\
  }\href {\doibase 10.1109/jstqe.2012.2195159} {\bibfield  {journal} {\bibinfo
  {journal} {IEEE J. Select. Topics Quantum Electron.}\ }\textbf {\bibinfo
  {volume} {18}},\ \bibinfo {pages} {1733} (\bibinfo {year}
  {2012})}\BibitemShut {NoStop}%
\bibitem [{\citenamefont {Dousse}\ \emph {et~al.}(2008)\citenamefont {Dousse},
  \citenamefont {Lanco}, \citenamefont {Suffczynski}, \citenamefont {Semenova},
  \citenamefont {Miard}, \citenamefont {Lemaitre}, \citenamefont {Sagnes},
  \citenamefont {Roblin}, \citenamefont {Bloch},\ and\ \citenamefont
  {Senellart}}]{Dousse.2008}%
  \BibitemOpen
  \bibfield  {author} {\bibinfo {author} {\bibfnamefont {A.}~\bibnamefont
  {Dousse}}, \bibinfo {author} {\bibfnamefont {L.}~\bibnamefont {Lanco}},
  \bibinfo {author} {\bibfnamefont {J.}~\bibnamefont {Suffczynski}}, \bibinfo
  {author} {\bibfnamefont {E.}~\bibnamefont {Semenova}}, \bibinfo {author}
  {\bibfnamefont {A.}~\bibnamefont {Miard}}, \bibinfo {author} {\bibfnamefont
  {A.}~\bibnamefont {Lemaitre}}, \bibinfo {author} {\bibfnamefont
  {I.}~\bibnamefont {Sagnes}}, \bibinfo {author} {\bibfnamefont
  {C.}~\bibnamefont {Roblin}}, \bibinfo {author} {\bibfnamefont
  {J.}~\bibnamefont {Bloch}}, \ and\ \bibinfo {author} {\bibfnamefont
  {P.}~\bibnamefont {Senellart}},\ }\href {\doibase
  10.1103/PhysRevLett.101.267404} {\bibfield  {journal} {\bibinfo  {journal}
  {Phys. Rev. Lett.}\ }\textbf {\bibinfo {volume} {101}},\ \bibinfo {pages}
  {267404} (\bibinfo {year} {2008})}\BibitemShut {NoStop}%
\bibitem [{\citenamefont {Gschrey}\ \emph {et~al.}(2013)\citenamefont
  {Gschrey}, \citenamefont {Gericke}, \citenamefont {Sch{\"u}{\ss}ler},
  \citenamefont {Schmidt}, \citenamefont {Schulze}, \citenamefont {Heindel},
  \citenamefont {Rodt}, \citenamefont {Strittmatter},\ and\ \citenamefont
  {Reitzenstein}}]{Gschrey.2013}%
  \BibitemOpen
  \bibfield  {author} {\bibinfo {author} {\bibfnamefont {M.}~\bibnamefont
  {Gschrey}}, \bibinfo {author} {\bibfnamefont {F.}~\bibnamefont {Gericke}},
  \bibinfo {author} {\bibfnamefont {A.}~\bibnamefont {Sch{\"u}{\ss}ler}},
  \bibinfo {author} {\bibfnamefont {R.}~\bibnamefont {Schmidt}}, \bibinfo
  {author} {\bibfnamefont {J.-H.}\ \bibnamefont {Schulze}}, \bibinfo {author}
  {\bibfnamefont {T.}~\bibnamefont {Heindel}}, \bibinfo {author} {\bibfnamefont
  {S.}~\bibnamefont {Rodt}}, \bibinfo {author} {\bibfnamefont {A.}~\bibnamefont
  {Strittmatter}}, \ and\ \bibinfo {author} {\bibfnamefont {S.}~\bibnamefont
  {Reitzenstein}},\ }\href {\doibase 10.1063/1.4812343} {\bibfield  {journal}
  {\bibinfo  {journal} {Appl. Phys. Lett.}\ }\textbf {\bibinfo {volume}
  {102}},\ \bibinfo {pages} {251113} (\bibinfo {year} {2013})}\BibitemShut
  {NoStop}%
\bibitem [{\citenamefont {{M. Bayer}}\ and\ \citenamefont {{A.
  Forchel}}(2002)}]{M.Bayer.2002}%
  \BibitemOpen
  \bibfield  {author} {\bibinfo {author} {\bibnamefont {{M. Bayer}}}\ and\
  \bibinfo {author} {\bibnamefont {{A. Forchel}}},\ }\href {\doibase
  10.1103/physrevb.65.041308} {\bibfield  {journal} {\bibinfo  {journal} {Phys.
  Rev. B}\ }\textbf {\bibinfo {volume} {65}},\ \bibinfo {pages} {041308}
  (\bibinfo {year} {2002})}\BibitemShut {NoStop}%
\bibitem [{\citenamefont {{Y. Chen}}(2015)}]{Y.Chen.2015}%
  \BibitemOpen
  \bibfield  {author} {\bibinfo {author} {\bibnamefont {{Y. Chen}}},\ }\href
  {\doibase 10.1016/j.mee.2015.02.042} {\bibfield  {journal} {\bibinfo
  {journal} {Microelectronic Engineering}\ }\textbf {\bibinfo {volume} {135}},\
  \bibinfo {pages} {57} (\bibinfo {year} {2015})}\BibitemShut {NoStop}%
\bibitem [{\citenamefont {{V. Kudryashov}}\ \emph {et~al.}(2003)\citenamefont
  {{V. Kudryashov}}, \citenamefont {{X.-C. Yuan}}, \citenamefont {{W.-C.
  Cheong}},\ and\ \citenamefont {{K. Radhakrishnan}}}]{V.Kudryashov.2003}%
  \BibitemOpen
  \bibfield  {author} {\bibinfo {author} {\bibnamefont {{V. Kudryashov}}},
  \bibinfo {author} {\bibnamefont {{X.-C. Yuan}}}, \bibinfo {author}
  {\bibnamefont {{W.-C. Cheong}}}, \ and\ \bibinfo {author} {\bibnamefont {{K.
  Radhakrishnan}}},\ }\href {\doibase 10.1016/s0167-9317(03)00083-2} {\bibfield
   {journal} {\bibinfo  {journal} {Microelectronic Engineering}\ }\textbf
  {\bibinfo {volume} {67-68}},\ \bibinfo {pages} {306} (\bibinfo {year}
  {2003})}\BibitemShut {NoStop}%
\bibitem [{\citenamefont {{W. H. Teh}}\ and\ \citenamefont {{C. G.
  Smith}}(2003)}]{W.H.Teh.2003}%
  \BibitemOpen
  \bibfield  {author} {\bibinfo {author} {\bibnamefont {{W. H. Teh}}}\ and\
  \bibinfo {author} {\bibnamefont {{C. G. Smith}}},\ }\href {\doibase
  10.1116/1.1629290} {\bibfield  {journal} {\bibinfo  {journal} {J. Vac. Sci.
  Technol. B}\ }\textbf {\bibinfo {volume} {21}},\ \bibinfo {pages} {3007}
  (\bibinfo {year} {2003})}\BibitemShut {NoStop}%
\bibitem [{\citenamefont {Schnauber}\ \emph {et~al.}(2016)\citenamefont
  {Schnauber}, \citenamefont {Schmidt}, \citenamefont {Kaganskiy},
  \citenamefont {Heuser}, \citenamefont {Gschrey}, \citenamefont {Rodt},\ and\
  \citenamefont {Reitzenstein}}]{Schnauber.2016}%
  \BibitemOpen
  \bibfield  {author} {\bibinfo {author} {\bibfnamefont {P.}~\bibnamefont
  {Schnauber}}, \bibinfo {author} {\bibfnamefont {R.}~\bibnamefont {Schmidt}},
  \bibinfo {author} {\bibfnamefont {A.}~\bibnamefont {Kaganskiy}}, \bibinfo
  {author} {\bibfnamefont {T.}~\bibnamefont {Heuser}}, \bibinfo {author}
  {\bibfnamefont {M.}~\bibnamefont {Gschrey}}, \bibinfo {author} {\bibfnamefont
  {S.}~\bibnamefont {Rodt}}, \ and\ \bibinfo {author} {\bibfnamefont
  {S.}~\bibnamefont {Reitzenstein}},\ }\href {\doibase
  10.1088/0957-4484/27/19/195301} {\bibfield  {journal} {\bibinfo  {journal}
  {Nanotechnology}\ }\textbf {\bibinfo {volume} {27}},\ \bibinfo {pages}
  {195301} (\bibinfo {year} {2016})}\BibitemShut {NoStop}%
\bibitem [{\citenamefont {{M. Gschrey}}\ \emph {et~al.}(2014)\citenamefont {{M.
  Gschrey}}, \citenamefont {{R. Schmidt}}, \citenamefont {{A. Kaganskiy}},
  \citenamefont {{S. Rodt}},\ and\ \citenamefont {{S.
  Reitzenstein}}}]{M.Gschrey.2014}%
  \BibitemOpen
  \bibfield  {author} {\bibinfo {author} {\bibnamefont {{M. Gschrey}}},
  \bibinfo {author} {\bibnamefont {{R. Schmidt}}}, \bibinfo {author}
  {\bibnamefont {{A. Kaganskiy}}}, \bibinfo {author} {\bibnamefont {{S.
  Rodt}}}, \ and\ \bibinfo {author} {\bibnamefont {{S. Reitzenstein}}},\ }\href
  {\doibase 10.1116/1.4896671} {\bibfield  {journal} {\bibinfo  {journal} {J.
  Vac. Sci. Technol. B}\ }\textbf {\bibinfo {volume} {32}},\ \bibinfo {pages}
  {061601} (\bibinfo {year} {2014})}\BibitemShut {NoStop}%
\bibitem [{\citenamefont {{W.-M. Yeh}}\ \emph {et~al.}(2010)\citenamefont
  {{W.-M. Yeh}}, \citenamefont {{D. E. Noga}}, \citenamefont {{R. A. Lawson}},
  \citenamefont {{L. M. Tolbert}},\ and\ \citenamefont {{C. L.
  Henderson}}}]{W.M.Yeh.2010}%
  \BibitemOpen
  \bibfield  {author} {\bibinfo {author} {\bibnamefont {{W.-M. Yeh}}}, \bibinfo
  {author} {\bibnamefont {{D. E. Noga}}}, \bibinfo {author} {\bibnamefont {{R.
  A. Lawson}}}, \bibinfo {author} {\bibnamefont {{L. M. Tolbert}}}, \ and\
  \bibinfo {author} {\bibnamefont {{C. L. Henderson}}},\ }\href {\doibase
  10.1116/1.3518136} {\bibfield  {journal} {\bibinfo  {journal} {J. Vac. Sci.
  Technol. B}\ }\textbf {\bibinfo {volume} {28}},\ \bibinfo {pages} {C6S6}
  (\bibinfo {year} {2010})}\BibitemShut {NoStop}%
\bibitem [{\citenamefont {{L. Ressier}}\ \emph {et~al.}(2007)\citenamefont {{L.
  Ressier}}, \citenamefont {{J. Grisolia}}, \citenamefont {{C. Martin}},
  \citenamefont {{J. P. Peyrade}}, \citenamefont {{B. Viallet}},\ and\
  \citenamefont {{C. Vieu}}}]{L.Ressier.2007}%
  \BibitemOpen
  \bibfield  {author} {\bibinfo {author} {\bibnamefont {{L. Ressier}}},
  \bibinfo {author} {\bibnamefont {{J. Grisolia}}}, \bibinfo {author}
  {\bibnamefont {{C. Martin}}}, \bibinfo {author} {\bibnamefont {{J. P.
  Peyrade}}}, \bibinfo {author} {\bibnamefont {{B. Viallet}}}, \ and\ \bibinfo
  {author} {\bibnamefont {{C. Vieu}}},\ }\href {\doibase
  10.1016/j.ultramic.2007.04.013} {\bibfield  {journal} {\bibinfo  {journal}
  {Ultramicroscopy}\ }\textbf {\bibinfo {volume} {107}},\ \bibinfo {pages}
  {985} (\bibinfo {year} {2007})}\BibitemShut {NoStop}%
\bibitem [{\citenamefont {{I. Zailer}}\ \emph {et~al.}(1996)\citenamefont {{I.
  Zailer}}, \citenamefont {{J. E. F. Frost}}, \citenamefont {{V.
  Chabasseur-Molyneux}}, \citenamefont {{C. J. B. Ford}},\ and\ \citenamefont
  {{M. Pepper}}}]{I.Zailer.1996}%
  \BibitemOpen
  \bibfield  {author} {\bibinfo {author} {\bibnamefont {{I. Zailer}}}, \bibinfo
  {author} {\bibnamefont {{J. E. F. Frost}}}, \bibinfo {author} {\bibnamefont
  {{V. Chabasseur-Molyneux}}}, \bibinfo {author} {\bibnamefont {{C. J. B.
  Ford}}}, \ and\ \bibinfo {author} {\bibnamefont {{M. Pepper}}},\ }\href
  {\doibase 10.1088/0268-1242/11/8/021} {\bibfield  {journal} {\bibinfo
  {journal} {Semicond. Sci. Technol.}\ }\textbf {\bibinfo {volume} {11}},\
  \bibinfo {pages} {1235} (\bibinfo {year} {1996})}\BibitemShut {NoStop}%
\bibitem [{\citenamefont {{A. C. F. Hoole}}\ \emph {et~al.}(1997)\citenamefont
  {{A. C. F. Hoole}}, \citenamefont {{M. E. Welland}},\ and\ \citenamefont {{A.
  N. Broers}}}]{A.C.F.Hoole.1997}%
  \BibitemOpen
  \bibfield  {author} {\bibinfo {author} {\bibnamefont {{A. C. F. Hoole}}},
  \bibinfo {author} {\bibnamefont {{M. E. Welland}}}, \ and\ \bibinfo {author}
  {\bibnamefont {{A. N. Broers}}},\ }\href {\doibase
  10.1088/0268-1242/12/9/017} {\bibfield  {journal} {\bibinfo  {journal}
  {Semicond. Sci. Technol.}\ }\textbf {\bibinfo {volume} {12}},\ \bibinfo
  {pages} {1166} (\bibinfo {year} {1997})}\BibitemShut {NoStop}%
\bibitem [{\citenamefont {Kaganskiy}\ \emph {et~al.}(2015)\citenamefont
  {Kaganskiy}, \citenamefont {Gschrey}, \citenamefont {Schlehahn},
  \citenamefont {Schmidt}, \citenamefont {Schulze}, \citenamefont {Heindel},
  \citenamefont {Strittmatter}, \citenamefont {Rodt},\ and\ \citenamefont
  {Reitzenstein}}]{Kaganskiy.2015}%
  \BibitemOpen
  \bibfield  {author} {\bibinfo {author} {\bibfnamefont {A.}~\bibnamefont
  {Kaganskiy}}, \bibinfo {author} {\bibfnamefont {M.}~\bibnamefont {Gschrey}},
  \bibinfo {author} {\bibfnamefont {A.}~\bibnamefont {Schlehahn}}, \bibinfo
  {author} {\bibfnamefont {R.}~\bibnamefont {Schmidt}}, \bibinfo {author}
  {\bibfnamefont {J.-H.}\ \bibnamefont {Schulze}}, \bibinfo {author}
  {\bibfnamefont {T.}~\bibnamefont {Heindel}}, \bibinfo {author} {\bibfnamefont
  {A.}~\bibnamefont {Strittmatter}}, \bibinfo {author} {\bibfnamefont
  {S.}~\bibnamefont {Rodt}}, \ and\ \bibinfo {author} {\bibfnamefont
  {S.}~\bibnamefont {Reitzenstein}},\ }\href {\doibase 10.1063/1.4926995}
  {\bibfield  {journal} {\bibinfo  {journal} {Rev. Sci. Instrum.}\ }\textbf
  {\bibinfo {volume} {86}},\ \bibinfo {pages} {073903} (\bibinfo {year}
  {2015})}\BibitemShut {NoStop}%
\bibitem [{\citenamefont {Gschrey}\ \emph
  {et~al.}(2015{\natexlab{a}})\citenamefont {Gschrey}, \citenamefont {Thoma},
  \citenamefont {Schnauber}, \citenamefont {Seifried}, \citenamefont {Schmidt},
  \citenamefont {Wohlfeil}, \citenamefont {Kruger}, \citenamefont {Schulze},
  \citenamefont {Heindel}, \citenamefont {Burger}, \citenamefont {Schmidt},
  \citenamefont {Strittmatter}, \citenamefont {Rodt},\ and\ \citenamefont
  {Reitzenstein}}]{Gschrey.2015}%
  \BibitemOpen
  \bibfield  {author} {\bibinfo {author} {\bibfnamefont {M.}~\bibnamefont
  {Gschrey}}, \bibinfo {author} {\bibfnamefont {A.}~\bibnamefont {Thoma}},
  \bibinfo {author} {\bibfnamefont {P.}~\bibnamefont {Schnauber}}, \bibinfo
  {author} {\bibfnamefont {M.}~\bibnamefont {Seifried}}, \bibinfo {author}
  {\bibfnamefont {R.}~\bibnamefont {Schmidt}}, \bibinfo {author} {\bibfnamefont
  {B.}~\bibnamefont {Wohlfeil}}, \bibinfo {author} {\bibfnamefont
  {L.}~\bibnamefont {Kruger}}, \bibinfo {author} {\bibfnamefont {J.-H.}\
  \bibnamefont {Schulze}}, \bibinfo {author} {\bibfnamefont {T.}~\bibnamefont
  {Heindel}}, \bibinfo {author} {\bibfnamefont {S.}~\bibnamefont {Burger}},
  \bibinfo {author} {\bibfnamefont {F.}~\bibnamefont {Schmidt}}, \bibinfo
  {author} {\bibfnamefont {A.}~\bibnamefont {Strittmatter}}, \bibinfo {author}
  {\bibfnamefont {S.}~\bibnamefont {Rodt}}, \ and\ \bibinfo {author}
  {\bibfnamefont {S.}~\bibnamefont {Reitzenstein}},\ }\href {\doibase
  10.1038/ncomms8662} {\bibfield  {journal} {\bibinfo  {journal} {Nat.
  Commun.}\ }\textbf {\bibinfo {volume} {6}},\ \bibinfo {pages} {7662}
  (\bibinfo {year} {2015}{\natexlab{a}})}\BibitemShut {NoStop}%
\bibitem [{\citenamefont {{M. Davanco}}\ \emph {et~al.}(2011)\citenamefont {{M.
  Davanco}}, \citenamefont {{M. T. Rakher}}, \citenamefont {{D. Schuh}},
  \citenamefont {{A. Badolato}},\ and\ \citenamefont {{K.
  Srinivasan}}}]{M.Davanco.2011}%
  \BibitemOpen
  \bibfield  {author} {\bibinfo {author} {\bibnamefont {{M. Davanco}}},
  \bibinfo {author} {\bibnamefont {{M. T. Rakher}}}, \bibinfo {author}
  {\bibnamefont {{D. Schuh}}}, \bibinfo {author} {\bibnamefont {{A.
  Badolato}}}, \ and\ \bibinfo {author} {\bibnamefont {{K. Srinivasan}}},\
  }\href {\doibase 10.1063/1.3615051} {\bibfield  {journal} {\bibinfo
  {journal} {Appl. Phys. Lett.}\ }\textbf {\bibinfo {volume} {99}},\ \bibinfo
  {pages} {041102} (\bibinfo {year} {2011})}\BibitemShut {NoStop}%
\bibitem [{\citenamefont {{T. D. Happ}}\ \emph {et~al.}(2002)\citenamefont {{T.
  D. Happ}}, \citenamefont {{I. I. Tartakovskii}}, \citenamefont {{V. D.
  Kulakovskii}}, \citenamefont {{J.-P. Reithmaier}}, \citenamefont {{M.
  Kamp}},\ and\ \citenamefont {{A. Forchel}}}]{T.D.Happ.2002}%
  \BibitemOpen
  \bibfield  {author} {\bibinfo {author} {\bibnamefont {{T. D. Happ}}},
  \bibinfo {author} {\bibnamefont {{I. I. Tartakovskii}}}, \bibinfo {author}
  {\bibnamefont {{V. D. Kulakovskii}}}, \bibinfo {author} {\bibnamefont {{J.-P.
  Reithmaier}}}, \bibinfo {author} {\bibnamefont {{M. Kamp}}}, \ and\ \bibinfo
  {author} {\bibnamefont {{A. Forchel}}},\ }\href {\doibase
  10.1103/physrevb.66.041303} {\bibfield  {journal} {\bibinfo  {journal} {Phys.
  Rev. B}\ }\textbf {\bibinfo {volume} {66}},\ \bibinfo {pages} {041303}
  (\bibinfo {year} {2002})}\BibitemShut {NoStop}%
\bibitem [{\citenamefont {{J. Zhang}}\ \emph {et~al.}(2015)\citenamefont {{J.
  Zhang}}, \citenamefont {{K. Cao}}, \citenamefont {{X. S. Wang}},\ and\
  \citenamefont {{B. Cui}}}]{J.Zhang.2015}%
  \BibitemOpen
  \bibfield  {author} {\bibinfo {author} {\bibnamefont {{J. Zhang}}}, \bibinfo
  {author} {\bibnamefont {{K. Cao}}}, \bibinfo {author} {\bibnamefont {{X. S.
  Wang}}}, \ and\ \bibinfo {author} {\bibnamefont {{B. Cui}}},\ }\href
  {\doibase 10.1039/c5cc07117h} {\bibfield  {journal} {\bibinfo  {journal}
  {Chem. Commun.}\ }\textbf {\bibinfo {volume} {51}},\ \bibinfo {pages} {17592}
  (\bibinfo {year} {2015})}\BibitemShut {NoStop}%
\bibitem [{\citenamefont {{G. Karbasian}}\ \emph {et~al.}(2012)\citenamefont
  {{G. Karbasian}}, \citenamefont {{P. J. Fay}}, \citenamefont {{H. Xing}},
  \citenamefont {{D. Jena}}, \citenamefont {{A. O. Orlov}},\ and\ \citenamefont
  {{G. L. Snider}}}]{G.Karbasian.2012}%
  \BibitemOpen
  \bibfield  {author} {\bibinfo {author} {\bibnamefont {{G. Karbasian}}},
  \bibinfo {author} {\bibnamefont {{P. J. Fay}}}, \bibinfo {author}
  {\bibnamefont {{H. Xing}}}, \bibinfo {author} {\bibnamefont {{D. Jena}}},
  \bibinfo {author} {\bibnamefont {{A. O. Orlov}}}, \ and\ \bibinfo {author}
  {\bibnamefont {{G. L. Snider}}},\ }\href {\doibase 10.1116/1.4750217}
  {\bibfield  {journal} {\bibinfo  {journal} {J. Vac. Sci. Technol. B}\
  }\textbf {\bibinfo {volume} {30}},\ \bibinfo {pages} {06FI01} (\bibinfo
  {year} {2012})}\BibitemShut {NoStop}%
\bibitem [{\citenamefont {{L. Brigo}}\ \emph {et~al.}(2012)\citenamefont {{L.
  Brigo}}, \citenamefont {{V. Auzelyte}}, \citenamefont {{K. A. Lister}},
  \citenamefont {{J. Brugger}},\ and\ \citenamefont {{G.
  Brusatin}}}]{L.Brigo.2012}%
  \BibitemOpen
  \bibfield  {author} {\bibinfo {author} {\bibnamefont {{L. Brigo}}}, \bibinfo
  {author} {\bibnamefont {{V. Auzelyte}}}, \bibinfo {author} {\bibnamefont {{K.
  A. Lister}}}, \bibinfo {author} {\bibnamefont {{J. Brugger}}}, \ and\
  \bibinfo {author} {\bibnamefont {{G. Brusatin}}},\ }\href {\doibase
  10.1088/0957-4484/23/32/325302} {\bibfield  {journal} {\bibinfo  {journal}
  {Nanotechnology}\ }\textbf {\bibinfo {volume} {23}},\ \bibinfo {pages}
  {325302} (\bibinfo {year} {2012})}\BibitemShut {NoStop}%
\bibitem [{\citenamefont {{A. P. G. Robinson}}\ \emph
  {et~al.}(1999)\citenamefont {{A. P. G. Robinson}}, \citenamefont {{R. E.
  Palmer}}, \citenamefont {{T. Tada}}, \citenamefont {{T. Kanayama}},
  \citenamefont {{M. T. Allen}}, \citenamefont {{J. A. Preece}},\ and\
  \citenamefont {{K. D. M. Harris}}}]{A.P.G.Robinson.1999}%
  \BibitemOpen
  \bibfield  {author} {\bibinfo {author} {\bibnamefont {{A. P. G. Robinson}}},
  \bibinfo {author} {\bibnamefont {{R. E. Palmer}}}, \bibinfo {author}
  {\bibnamefont {{T. Tada}}}, \bibinfo {author} {\bibnamefont {{T. Kanayama}}},
  \bibinfo {author} {\bibnamefont {{M. T. Allen}}}, \bibinfo {author}
  {\bibnamefont {{J. A. Preece}}}, \ and\ \bibinfo {author} {\bibnamefont {{K.
  D. M. Harris}}},\ }\href {\doibase 10.1088/0022-3727/32/16/102} {\bibfield
  {journal} {\bibinfo  {journal} {J. Phys. D: Appl. Phys.}\ }\textbf {\bibinfo
  {volume} {32}},\ \bibinfo {pages} {L75} (\bibinfo {year} {1999})}\BibitemShut
  {NoStop}%
\bibitem [{\citenamefont {Oyama}\ \emph {et~al.}(2011)\citenamefont {Oyama},
  \citenamefont {Oshima}, \citenamefont {Yamamoto}, \citenamefont {Tagawa},\
  and\ \citenamefont {Washio}}]{Oyama.2011}%
  \BibitemOpen
  \bibfield  {author} {\bibinfo {author} {\bibfnamefont {T.~G.}\ \bibnamefont
  {Oyama}}, \bibinfo {author} {\bibfnamefont {A.}~\bibnamefont {Oshima}},
  \bibinfo {author} {\bibfnamefont {H.}~\bibnamefont {Yamamoto}}, \bibinfo
  {author} {\bibfnamefont {S.}~\bibnamefont {Tagawa}}, \ and\ \bibinfo {author}
  {\bibfnamefont {M.}~\bibnamefont {Washio}},\ }\href {\doibase
  10.1143/APEX.4.076501} {\bibfield  {journal} {\bibinfo  {journal} {Appl.
  Phys. Express}\ }\textbf {\bibinfo {volume} {4}},\ \bibinfo {pages} {076501}
  (\bibinfo {year} {2011})}\BibitemShut {NoStop}%
\bibitem [{\citenamefont {{M. A. Mohammad}}\ \emph {et~al.}(2012)\citenamefont
  {{M. A. Mohammad}}, \citenamefont {{K. Koshelev}}, \citenamefont {{T. Fito}},
  \citenamefont {{D. A. Z. Zheng}}, \citenamefont {{M. Stepanova}},\ and\
  \citenamefont {{S. Dew}}}]{M.A.Mohammad.2012}%
  \BibitemOpen
  \bibfield  {author} {\bibinfo {author} {\bibnamefont {{M. A. Mohammad}}},
  \bibinfo {author} {\bibnamefont {{K. Koshelev}}}, \bibinfo {author}
  {\bibnamefont {{T. Fito}}}, \bibinfo {author} {\bibnamefont {{D. A. Z.
  Zheng}}}, \bibinfo {author} {\bibnamefont {{M. Stepanova}}}, \ and\ \bibinfo
  {author} {\bibnamefont {{S. Dew}}},\ }\href {\doibase 10.1143/jjap.51.06fc05}
  {\bibfield  {journal} {\bibinfo  {journal} {Jpn. J. Appl. Phys.}\ }\textbf
  {\bibinfo {volume} {51}},\ \bibinfo {pages} {06FC05} (\bibinfo {year}
  {2012})}\BibitemShut {NoStop}%
\bibitem [{\citenamefont {Gschrey}\ \emph
  {et~al.}(2015{\natexlab{b}})\citenamefont {Gschrey}, \citenamefont {Schmidt},
  \citenamefont {Schulze}, \citenamefont {Strittmatter}, \citenamefont {Rodt},\
  and\ \citenamefont {Reitzenstein}}]{Gschrey.2015b}%
  \BibitemOpen
  \bibfield  {author} {\bibinfo {author} {\bibfnamefont {M.}~\bibnamefont
  {Gschrey}}, \bibinfo {author} {\bibfnamefont {R.}~\bibnamefont {Schmidt}},
  \bibinfo {author} {\bibfnamefont {J.-H.}\ \bibnamefont {Schulze}}, \bibinfo
  {author} {\bibfnamefont {A.}~\bibnamefont {Strittmatter}}, \bibinfo {author}
  {\bibfnamefont {S.}~\bibnamefont {Rodt}}, \ and\ \bibinfo {author}
  {\bibfnamefont {S.}~\bibnamefont {Reitzenstein}},\ }\href {\doibase
  10.1116/1.4914914} {\bibfield  {journal} {\bibinfo  {journal} {J. Vac. Sci.
  Technol. B}\ }\textbf {\bibinfo {volume} {33}},\ \bibinfo {pages} {021603}
  (\bibinfo {year} {2015}{\natexlab{b}})}\BibitemShut {NoStop}%
\bibitem [{\citenamefont {Wuhrer}\ and\ \citenamefont
  {Moran}(2016)}]{Wuhrer.2016}%
  \BibitemOpen
  \bibfield  {author} {\bibinfo {author} {\bibfnamefont {R.}~\bibnamefont
  {Wuhrer}}\ and\ \bibinfo {author} {\bibfnamefont {K.}~\bibnamefont {Moran}},\
  }\href {\doibase 10.1088/1757-899X/109/1/012019} {\bibfield  {journal}
  {\bibinfo  {journal} {IOP Conf. Ser.: Mater. Sci. Eng.}\ }\textbf {\bibinfo
  {volume} {109}},\ \bibinfo {pages} {012019} (\bibinfo {year}
  {2016})}\BibitemShut {NoStop}%
\bibitem [{\citenamefont {Koshelev}\ \emph {et~al.}(2011)\citenamefont
  {Koshelev}, \citenamefont {Mohammad}, \citenamefont {Fito}, \citenamefont
  {Westra}, \citenamefont {Dew},\ and\ \citenamefont
  {Stepanova}}]{Koshelev.2011}%
  \BibitemOpen
  \bibfield  {author} {\bibinfo {author} {\bibfnamefont {K.}~\bibnamefont
  {Koshelev}}, \bibinfo {author} {\bibfnamefont {A.~M.}\ \bibnamefont
  {Mohammad}}, \bibinfo {author} {\bibfnamefont {T.}~\bibnamefont {Fito}},
  \bibinfo {author} {\bibfnamefont {K.~L.}\ \bibnamefont {Westra}}, \bibinfo
  {author} {\bibfnamefont {S.~K.}\ \bibnamefont {Dew}}, \ and\ \bibinfo
  {author} {\bibfnamefont {M.}~\bibnamefont {Stepanova}},\ }\href {\doibase
  10.1116/1.3640794} {\bibfield  {journal} {\bibinfo  {journal} {J. Vac. Sci.
  Technol. B}\ }\textbf {\bibinfo {volume} {29}},\ \bibinfo {pages} {06F306}
  (\bibinfo {year} {2011})}\BibitemShut {NoStop}%
\bibitem [{\citenamefont {Oyama}\ \emph {et~al.}(2012)\citenamefont {Oyama},
  \citenamefont {Enomoto}, \citenamefont {Hosaka}, \citenamefont {Oshima},
  \citenamefont {Washio},\ and\ \citenamefont {Tagawa}}]{Oyama.2012}%
  \BibitemOpen
  \bibfield  {author} {\bibinfo {author} {\bibfnamefont {T.~G.}\ \bibnamefont
  {Oyama}}, \bibinfo {author} {\bibfnamefont {K.}~\bibnamefont {Enomoto}},
  \bibinfo {author} {\bibfnamefont {Y.}~\bibnamefont {Hosaka}}, \bibinfo
  {author} {\bibfnamefont {A.}~\bibnamefont {Oshima}}, \bibinfo {author}
  {\bibfnamefont {M.}~\bibnamefont {Washio}}, \ and\ \bibinfo {author}
  {\bibfnamefont {S.}~\bibnamefont {Tagawa}},\ }\href {\doibase
  10.1143/APEX.5.036501} {\bibfield  {journal} {\bibinfo  {journal} {Appl.
  Phys. Express}\ }\textbf {\bibinfo {volume} {5}},\ \bibinfo {pages} {036501}
  (\bibinfo {year} {2012})}\BibitemShut {NoStop}%
\bibitem [{\citenamefont {Oyama}\ \emph {et~al.}(2014)\citenamefont {Oyama},
  \citenamefont {Nakamura}, \citenamefont {Oshima}, \citenamefont {Washio},\
  and\ \citenamefont {Tagawa}}]{Oyama.2014}%
  \BibitemOpen
  \bibfield  {author} {\bibinfo {author} {\bibfnamefont {T.~G.}\ \bibnamefont
  {Oyama}}, \bibinfo {author} {\bibfnamefont {H.}~\bibnamefont {Nakamura}},
  \bibinfo {author} {\bibfnamefont {A.}~\bibnamefont {Oshima}}, \bibinfo
  {author} {\bibfnamefont {M.}~\bibnamefont {Washio}}, \ and\ \bibinfo {author}
  {\bibfnamefont {S.}~\bibnamefont {Tagawa}},\ }\href {\doibase
  10.7567/APEX.7.036501} {\bibfield  {journal} {\bibinfo  {journal} {Appl.
  Phys. Express}\ }\textbf {\bibinfo {volume} {7}},\ \bibinfo {pages} {036501}
  (\bibinfo {year} {2014})}\BibitemShut {NoStop}%
\bibitem [{\citenamefont {Sidorkin}\ \emph {et~al.}(2008)\citenamefont
  {Sidorkin}, \citenamefont {{van der Drift}},\ and\ \citenamefont
  {Salemink}}]{Sidorkin.2008}%
  \BibitemOpen
  \bibfield  {author} {\bibinfo {author} {\bibfnamefont {V.}~\bibnamefont
  {Sidorkin}}, \bibinfo {author} {\bibfnamefont {E.}~\bibnamefont {{van der
  Drift}}}, \ and\ \bibinfo {author} {\bibfnamefont {H.}~\bibnamefont
  {Salemink}},\ }\href {\doibase 10.1116/1.2987965} {\bibfield  {journal}
  {\bibinfo  {journal} {J. Vac. Sci. Technol. B}\ }\textbf {\bibinfo {volume}
  {26}},\ \bibinfo {pages} {2049} (\bibinfo {year} {2008})}\BibitemShut
  {NoStop}%
\bibitem [{\citenamefont {Unai}\ \emph {et~al.}(2013)\citenamefont {Unai},
  \citenamefont {Puttaraksa}, \citenamefont {Pussadee}, \citenamefont
  {Singkarat}, \citenamefont {Rhodes}, \citenamefont {Whitlow},\ and\
  \citenamefont {Singkarat}}]{Unai.2013}%
  \BibitemOpen
  \bibfield  {author} {\bibinfo {author} {\bibfnamefont {S.}~\bibnamefont
  {Unai}}, \bibinfo {author} {\bibfnamefont {N.}~\bibnamefont {Puttaraksa}},
  \bibinfo {author} {\bibfnamefont {N.}~\bibnamefont {Pussadee}}, \bibinfo
  {author} {\bibfnamefont {K.}~\bibnamefont {Singkarat}}, \bibinfo {author}
  {\bibfnamefont {M.~W.}\ \bibnamefont {Rhodes}}, \bibinfo {author}
  {\bibfnamefont {H.~J.}\ \bibnamefont {Whitlow}}, \ and\ \bibinfo {author}
  {\bibfnamefont {S.}~\bibnamefont {Singkarat}},\ }\href {\doibase
  10.1016/j.mee.2012.05.010} {\bibfield  {journal} {\bibinfo  {journal}
  {Microelectron. Eng.}\ }\textbf {\bibinfo {volume} {102}},\ \bibinfo {pages}
  {18} (\bibinfo {year} {2013})}\BibitemShut {NoStop}%
\end{thebibliography}
\end{document}